\documentclass[aps,pra,a4paper,twocolumn,showpacs,longbibliography,showkeys,floatfix,superscriptaddress]{revtex4-1}

\usepackage{graphicx}
\usepackage{amsmath,amssymb,bm}
\usepackage{hyperref}

\usepackage{multirow}

\graphicspath{{images/}}

\newcommand{\figwidth}{0.98\columnwidth}

\newcommand{\bpcb}{\textnormal{({C}$_{5}$H$_{12}$N)$_{2}$CuBr$_{4}$}}

\newcommand{\spin}{\hat{S}}

\begin{document}
	\title{Anisotropy-Induced Soliton Excitation in Magnetized Strong-Rung Spin Ladders}
	
	\author{Yu. V. Krasnikova}
	\affiliation{P. L. Kapitza Institute for Physical Problems, RAS, Kosygina 2, 119334 Moscow, Russia}
	\email{krasnikova.mipt@gmail.com}

	\affiliation{Laboratory for Condensed Matter Physics, National Research University Higher School of Economics, Myasnitskaya str.20, 101000 Moscow, Russia}
	
	\author{S. C. Furuya}
	\affiliation{Condensed Matter Theory Laboratory, RIKEN, Wako, Saitama 351-0198, Japan}
	
	\author{V. N. Glazkov}
	\affiliation{P. L. Kapitza Institute for Physical Problems, RAS, Kosygina 2, 119334 Moscow, Russia}
	
	\affiliation{Laboratory for Condensed Matter Physics, National Research University Higher School of Economics, Myasnitskaya str.20, 101000 Moscow, Russia}
	
	\author{K. Yu. Povarov}
	\affiliation{Laboratory for Solid State Physics, ETH Z\"{u}rich, 8093 Z\"{u}rich, Switzerland}
	
	\author{D. Blosser}
	\affiliation{Laboratory for Solid State Physics, ETH Z\"{u}rich, 8093 Z\"{u}rich, Switzerland}
	
	\author{A. Zheludev}
	\affiliation{Laboratory for Solid State Physics, ETH Z\"{u}rich, 8093 Z\"{u}rich, Switzerland}
	
	\date{\today}
	
	\begin{abstract}
		We report low temperature electron spin resonance experimental and theoretical studies of an archetype $S=1/2$ strong-rung spin ladder material \bpcb. Unexpected dynamics is detected deep in the Tomonaga-Luttinger spin liquid regime. Close to the point where the system is half-magnetized (and believed to be equivalent to a gapless easy plane chain in zero field) we observed orientation-dependent spin gap and anomalous $g$-factor values. Field theoretical analysis demonstrates that the observed low-energy excitation modes in magnetized \bpcb\ are solitonic excitations caused by Dzyaloshinskii-Moriya interaction presence.
	\end{abstract}
	
	
	
	
	\maketitle

	\begin{figure}[h]
		\includegraphics[width=0.4\textwidth]{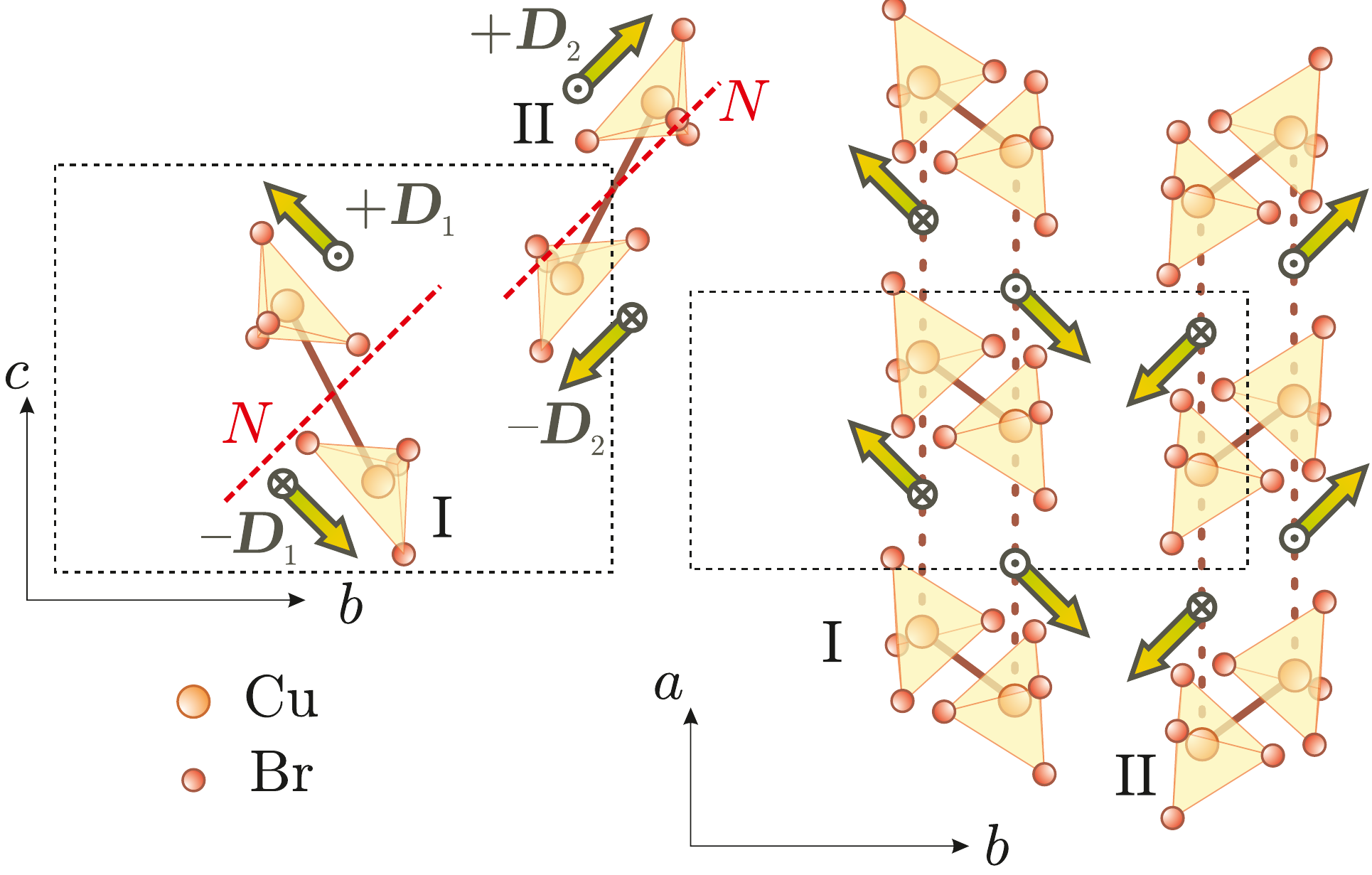}
		\caption{Schematic representation of BPCB crystal structure. I and II mark two types of ladders, $\textbf{D}_{1}, \textbf{D}_{2}$ - DM vectors directions, dashed line marks the $N$ direction as described in the text.}\label{fig:Geom01}
	\end{figure}

	\begin{figure}[h]
		\includegraphics[width=\figwidth]{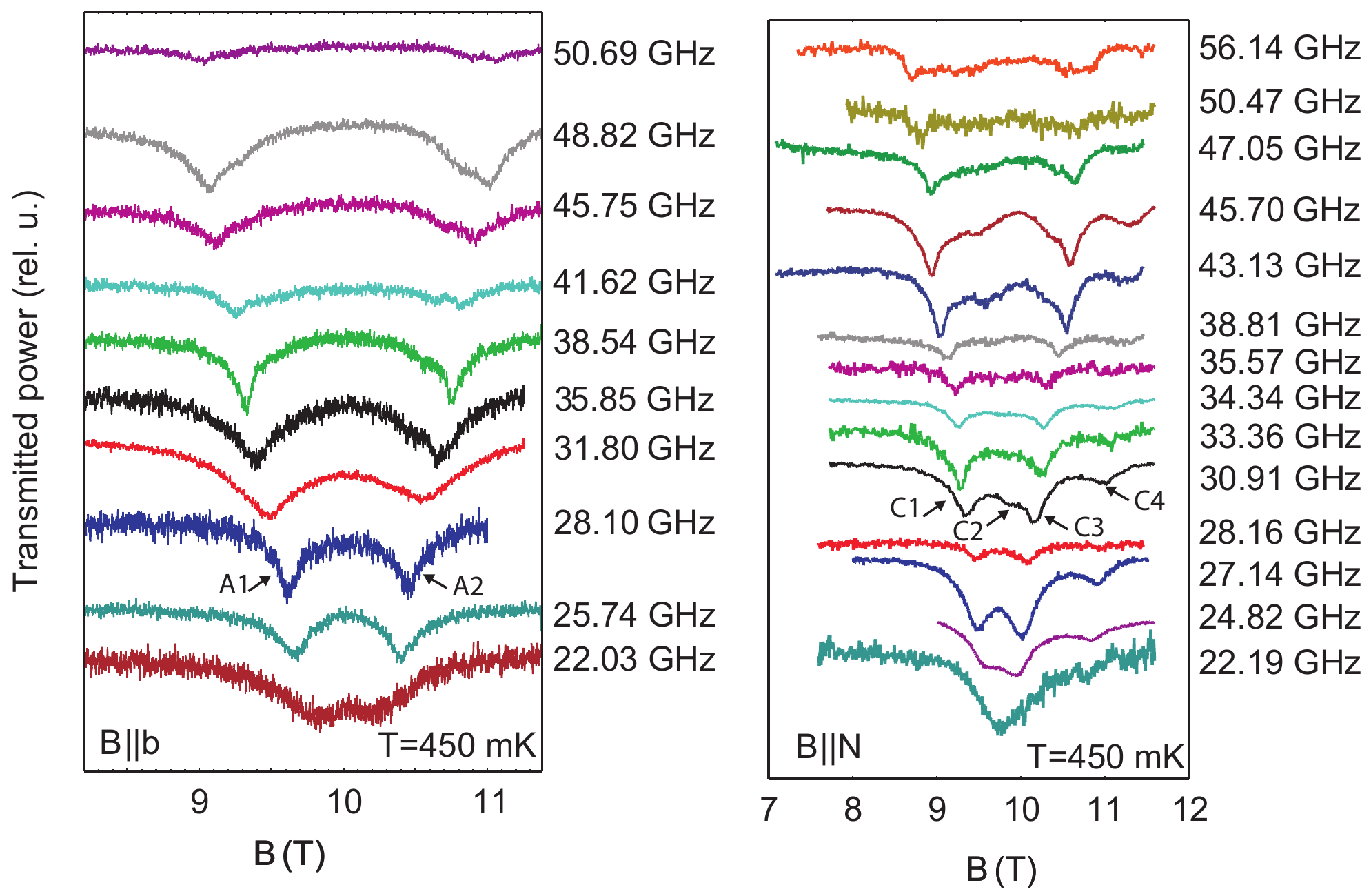}
		\caption{ESR absorption spectra at $T$=450 mK at different microwave frequencies. Left panel: magnetic field orientation $B\parallel b$. Right panel: magnetic field orientation $B\parallel N$.  A1, A2, C1, C2, C3, C4 mark different ESR absorption components for both orientations.}\label{fig:ESRlines}
	\end{figure}

	\begin{figure}[h]
		\includegraphics[width=\figwidth]{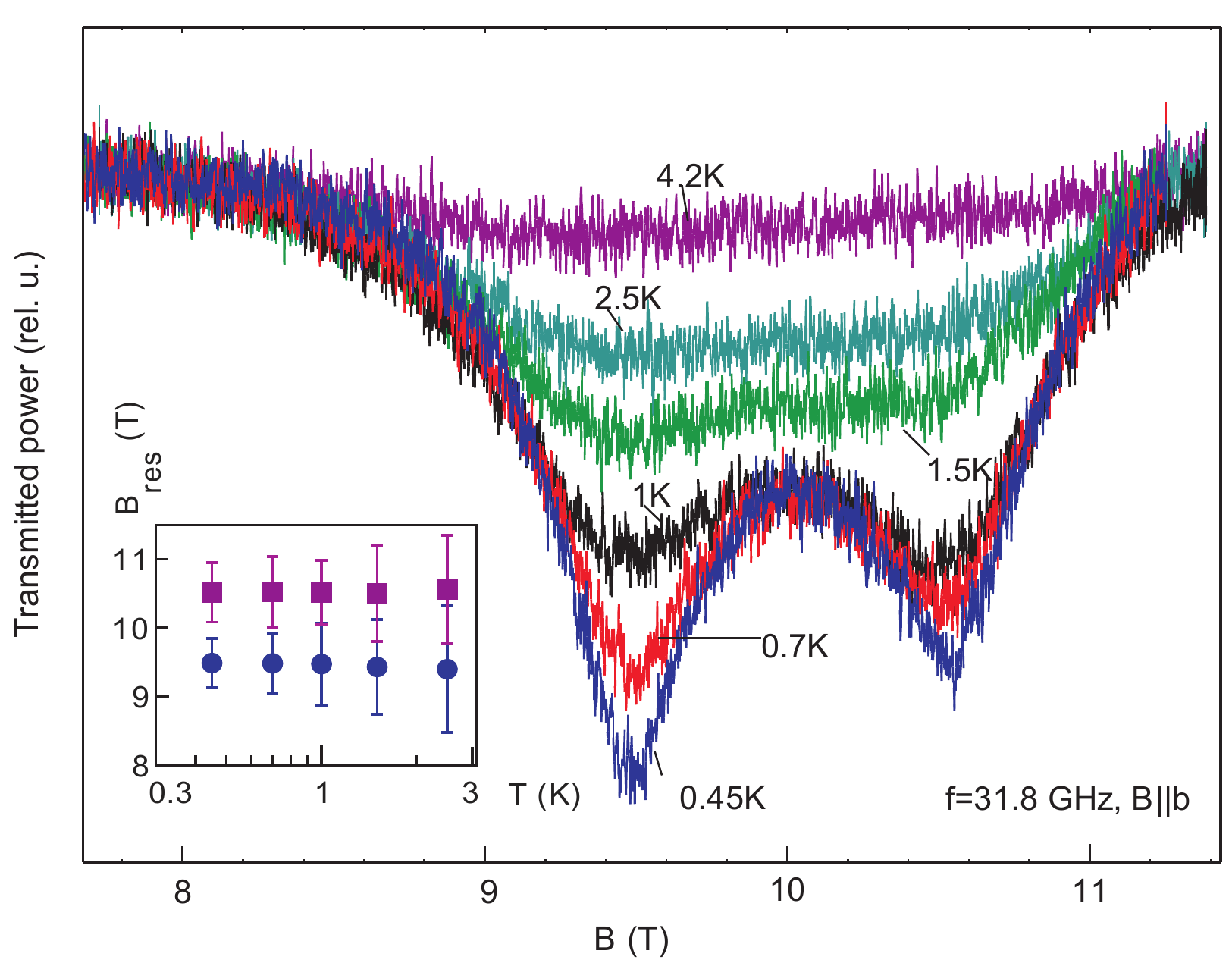}
		\caption{Temperature dependence of ESR absorption signal for $\nu$=31.8 GHz, sample orientation corresponds to $B\parallel b$. Insert: resonance field temperature dependence for both  ESR absorption components, circles -- left component, squares -- right component. Vertical bars show \emph{full} ESR linewidths at halfheight for both components.}\label{fig:tempdep}
	\end{figure}

	\begin{figure}[h]
		\includegraphics[width=0.5\textwidth]{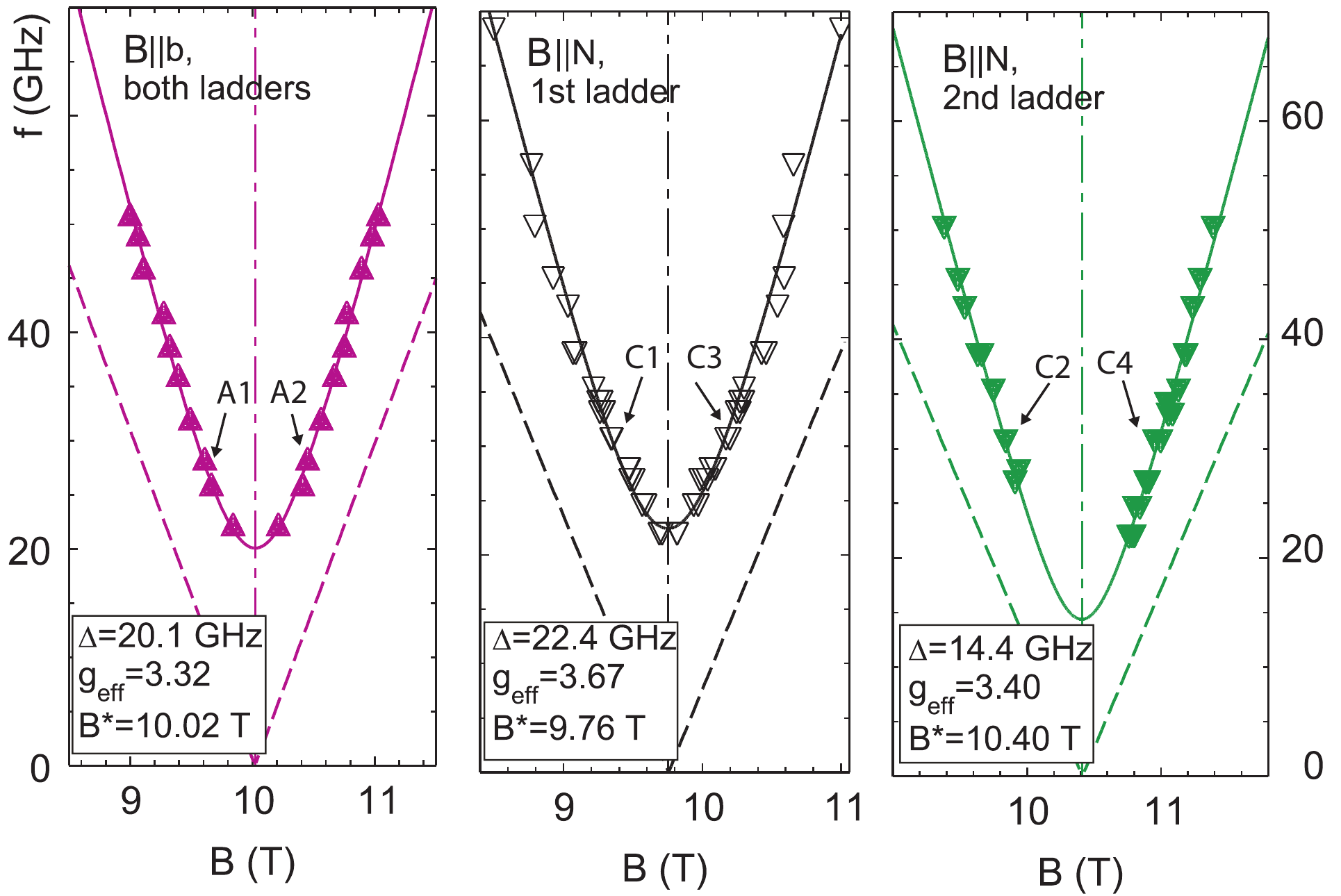}
		\caption{Frequency-field dependencies for both studied orientations $B\parallel N$ and $B\parallel b$ at $T$=450 mK. Symbols -- experimental data, solid curves -- fit by Eqn.~(\ref{eqn:fH}), dashed lines -- isotropic model with experimentally measured $g$-factors. A1, A2, C1, C2, C3, C4 mark different ESR absorption components for both orientations. The errorbars are within the symbol size. }\label{fig:freqfield}
	\end{figure}

	Understanging quantum many-body systems has always been a challenge. However, some exact solutions can be found for 1D systems~\cite{Giamarchi_1D}. There is a rich variety of low-dimensional magnets which could be approached in terms of Tomonaga-Luttinger spin liquid~\cite{Tomonaga_ProgThPhys_1950_TLL,*Luttinger_JMathPhys_1963_TLL,*LiebMatthis_JMathPhys_1965_TLL,Haldane_PRL_1980_TLSLconjecture,Giamarchi_exp}. One of the simplest and well described model for demonstrating the presence of Tomonaga-Luttinger spin liquid state (TLSL) is a two-leg spin ladder \cite{Jeong_PRL_2013_DIMPYnmr,*PovarovSchmidiger_PRB_2015_DimpyScaling,*JeongSchmidiger_PRL_2016_TLSLdichotomy,Thielemann_PRL_2009_BPCBtlsl,SM8}. Such a system has a gap in energy spectrum below the first critical field and is expected to have gapless spectrum in TLSL phase above the first critical field $B_{c1}$ up to the saturation field $B_{c2}$~\cite{Rice_EPL_1993_GapNoGapLadders,Dagotto_RepProgPhys_1999}. Moreover, mapping of the strong-rung spin ladder onto the exactly soluble XXZ spin chain can be realized in region between $B_{c1}$ and $B_{c2}$~\cite{Chaboussant_XXZ, Watson_PRL_2001_BPCBz2story, Bouillot_PRB_2011_bigBPCBpaper,GiamarchiTsvelik_PRB_1999_LadderBEC}.
	
	In the real systems anisotropic spin interactions are always present. They typically lead to significant low energy spectrum changes~\cite{Dender_benzoate,Glazkov_TlCuCl3,Zvyagin_CuPM,Povarov_Cs2CuCl4, Hikihara_bosonization}. Incorporating effects of such perturbations in TLSL description is one of the challenges in the research on 1D magnetic systems. However, not so many experimental methods allow to study the anisotropy induced effects in detail. Electron spin resonance spectroscopy (ESR) is perfectly suitable for this goal as it routinely allows energy resolution $\sim~0.005$ meV at $q=0$.
	
	In this Letter we report experimental and theoretical ESR studies for an archetype strong rung spin ladder model material \bpcb~(called BPCB)~\cite{Thielemann_PRB_2009_BPCBphasediagram,Thielemann_PRL_2009_BPCBtlsl, Bouillot_PRB_2011_bigBPCBpaper,Klanjsek_PSS_2010_BPCBnmr}.
	At low temperatures around the field corresponding to half-saturation of BPCB (i. e. deep in the TLSL regime) we found a gapped spectrum with unusual frequency-field dependencies. The conventional Heisenberg spin Hamiltonian of BPCB does not capture this behavior, as it predicts a complete softening of $q=0$ excitations near the half-saturation field. Field theoretical analysis shows that the symmetry allowed pattern of Dzyaloshinskii--Moriya interaction (which can be accompanied by symmetric anisotropic exchange interactions) results in formation of low energy solitons in the spectrum of BPCB. The predicted field dependencies match the experimental results well and the gap variation with the field direction fixes the orientation of Dzyaloshinskii--Moriya vectors.
	
	The material BPCB, crystallizing in a $P2_{1}/c$ monoclinic space group~\cite{Patyal_PRB_1990_BPCBxtal}, represents an almost ideal $S=1/2$ strong rung spin ladder model. The Cu$^{2+}$ magnetic ions form a ``rung'' $S=1/2$ dimer pairs with coupling $J_{\perp}\simeq 12.7$~K. Each dimer hosts an inversion center in the middle.  The dimers are in turn coupled into ladders by the interaction $J_{\parallel}\simeq 3.54$~K (Fig.~\ref{fig:Geom01}). In zero magnetic field this results in a quantum disordered singlet ground state with the gapped ``triplon'' $S=1$ magnetic excitations, the gap value being $\Delta_{\text{ZF}}\simeq 9.2$~K. Between the gap closing field $B_{c1}=6.6$~T and the saturation field $B_{c2}=13.6$ T~\cite{Watson_PRL_2001_BPCBz2story,Klanjsek_PSS_2010_BPCBnmr} the antiferromagnetic order exists below $T\simeq0.1$~K. Above this threshold temperature it is replaced by a strongly correlated TLSL state, persisting up to 1.5~K~\cite{Ruegg_PRL_2008_BPCBbulk}. Importantly, BPCB has two types of identical ladders oriented differently. They are equivalent for the field direction $B\parallel b$, and are most inequivalent for $B \parallel (bc)$ at 45$^\circ$ from $b$ axis~\cite{Blosser_BPCB,Zvyagin_BPCB}. This special orientation is denoted $B\parallel N$. Low symmetry of BPCB also allows anisotropic Dzyalloshinskii-Moriya (DM) interaction on the ladder legs. The DM vector $\mathbf{D}$ is uniform along the leg, is directed oppositely on two legs of the same ladder and is related by reflection symmetry in the adjacent ladders. No further constrains for its direction are present.
	
	The multifrequency ESR spectra have been recorded in the Kapitza Institute in a custom-made spectrometer equipped with $^{3}$He cryostat and 14~T cryomanget. The $m \approx 100$~mg deuterated BPCB crystal (from the same batch as in Refs.~\cite{BlosserBhartiya_PRL_2018_BPCBz2,Blosser_BPCB}) was placed in the cylindric microwave resonant cavity and the field dependent transmission spectra were recorded at the cavity eigenfrequencies. Two principal magnetic field directions, $B\parallel b$ and $B\parallel N$ were used. The data were collected well above the ordered state: from $T=0.45$~K to $20$~K in a wide field range.

	At $T>10$ K we observed normal paramagnetic resonance with the $g$-factor values 2.18 for $B \parallel b$, 2.04 and 2.29 for $B \parallel N$, which are close to the previously reported values~\cite{Patyal_PRB_1990_BPCBxtal, Zvyagin_BPCB}. On cooling below $T\simeq\Delta_{\text{ZF}}$ this signal looses intensity and splits, as expected for ESR of thermally activated triplet excitations \cite{Zvyagin_BPCB,Glazkov_PRB_2015_ESRDimpy, Glazkov_TlCuCl3, Glazkov_PHCC}.

	
	Besides this triplon-related low field absorption, we have detected a novel signal between two critical fields $B_{c1}$=6.6 T and $B_{c2}$=13.6 T at 450 mK (see Fig. \ref{fig:ESRlines}). This signal exists in a broad frequency range and looses intensity on heating (see Fig. \ref{fig:tempdep}). Therefore we can state that the observed modes are solely low-temperature phenomenon and they are connected with presence of TLSL phase in BPCB. We highlight that the temperature of our experiment is well above the ordering temperature. This allows to neglect interladder coupling in further analysis.

	Experimentally determined frequency-field diagrams are shown in Fig.~\ref{fig:freqfield}. They were fitted by equation:
	\begin{equation}
	2\pi\hbar\nu=\sqrt{{\Delta}^2 + \left(g_{\text{eff}}\mu_{B}(B - B^*)\right)^{2}},
	\label{eqn:fH}
	\end{equation}
	fit parameters are shown in Fig. \ref{fig:freqfield}. The gap $\Delta/(2\pi\hbar)$ depends on magnetic field orientation varying from $14$ to $22$~GHz for the orientations studied. Effective $g$-factor values vary from 3.2 to 3.7 for different field directions and are strongly renormalized compared to the bare values quoted above.

	Presence of the anisotropic gap and renormalization of effective $g$-factor are not captured by a simple Heisenberg Hamiltonian:
	\begin{align}\label{eq:LadderHamiltonian}\nonumber
	\mathcal{\hat{H}}=&\sum\limits_{j}J_{\perp}\hat{\mathbf{S}}_{j,1}\hat{\mathbf{S}}_{j,2}+J_{\parallel}(\hat{\mathbf{S}}_{j,1}\hat{\mathbf{S}}_{j+1,1}+\hat{\mathbf{S}}_{j,2}\hat{\mathbf{S}}_{j+1,2})\\
	&-g\mu_{B}\mathbf{B}(\mathbf{\hat{S}}_{j,1}+\mathbf{\hat{S}}_{j,2}),
	\end{align}
	with index $j$ runs along the ladder sites, and second index enumerates the ladder legs.

	For $J_{\perp}\gg J_{\parallel}$, all the low energy properties of ladder~(\ref{eq:LadderHamiltonian}) at $B > B_{c1}$ will essentially be defined by the two lowest energy states of the dimer on the rung of the ladder. This validates a ``ladder-to-chain'' mapping (see Ref.~\cite{Bouillot_PRB_2011_bigBPCBpaper} and references therein):
	\begin{equation}\label{EQ:LadderSpinMapping}
	\spin^{x,y}_{1}=-\frac{\hat{T}^{x,y}}{\sqrt{2}},~\spin^{x,y}_{2}=\frac{\hat{T}^{x,y}}{\sqrt{2}},~\spin^{z}_{1}=\spin^{z}_{2}=\frac{1+2\hat{T}^{z}}{4},
	\end{equation}
	Here, $\hat{T}^{\alpha}$ are the pseudospin operators that commute just like the normal spin operators. Assuming that $z$ is the field direction, the transformed Hamiltonian becomes the one of an easy plane pseudospin-1/2 chain:
	\begin{align}
	\label{EQ:HamiltonianXXZ}\nonumber
	\hat{\mathcal{H}}_{\text{XXZ}}=\sum\limits_{j}J_{\parallel}(\hat{T}_{j}^{x}\hat{T}_{j+1}^{x}+\hat{T}_{j}^{y}\hat{T}_{j+1}^{y}+\frac{1}{2}\hat{T}_{j}^{z}\hat{T}_{j+1}^{z})-\\
	-g\mu_{B}\biggl(B-\frac{J_{\perp}+J_{\parallel}/2}{g\mu_{B}}\biggr)\hat{T}^{z}_{j}
	\end{align}
	At a special magnetic field value $g\mu_{B}B^{\ast}=J_{\perp}+J_{\parallel}/2$ the system becomes equivalent to a non-magnetized easy plane chain. Thus, theoretical model \cite{Bouillot_PRB_2011_bigBPCBpaper,GiamarchiTsvelik_PRB_1999_LadderBEC} predicts softening of $q=0$ excitations at $B^{\ast}=(B_{c1}+B_{c2})/2$. Away from this point excitations spectrum should follow $2\pi\hbar\nu=g\mu_{B}|B-B^{\ast}|$.

	Our observations are clearly inconsistent with this idealized Heisenberg model. This urges us to consider the effect of peculiar type of Dzyaloshinskii--Moriya interaction (uniform along the leg, opposite on different legs) present in BPCB:
	\begin{equation}\label{eq:Hamilt_DM}
	\mathcal{\hat{H}}'=\sum\limits_{j}\mathbf{D}[\mathbf{\hat{S}}_{j,1}\times\mathbf{\hat{S}}_{j+1,1}]-\mathbf{D}[\mathbf{\hat{S}}_{j,2}\times\mathbf{\hat{S}}_{j+1,2}].
	\end{equation}
	The Dzyaloshiniskii--Moriya vector $\mathbf{D}=(D_{x},~0,~D_{z})$ may have both longitudinal and transverse components with respect to the external field. Under the transformation~(\ref{EQ:LadderSpinMapping}) the longitudinal part vanishes, and the transformed DM Hamiltonian becomes:
	\begin{equation}\label{eq:Hamilt_DM_XXZ}
	\hat{\mathcal{H}}_{\text{XXZ}}'=\frac{D_{x}}{\sqrt{2}}\sum_j\left(\hat{T}^{y}_{j}\hat{T}^{z}_{i+1}-\hat{T}^{z}_{j}\hat{T}^{y}_{j+1}\right).
	\end{equation}
	This means that our effective model is now the one of a spin chain with uniform DM interaction. Heisenberg spin chains with this type of interaction are known to demonstrate gapped ESR spectra at zero magnetic field \cite{Smirnov_DM_chain,Soldatov_DM,Povarov_Cs2CuCl4}. This would, in principle, explain the observed non-vanishing gap at field $B^{\ast}$. The present case, however, is \emph{strongly non-Heisenberg} in terms of pseudospin and the effective DM interaction is perpendicular to the field direction $z$.

	According to Kaplan--Shekhtman--Entin-Wohlman--Aharony (KSEA mechanism)~\cite{Kaplan_ZPB_1983_KSEA1,*Shekhtman_PRL_1992_KSEA2,*Shekhtman_PRB_1993_KSEA3, Zheludev_KSEA}, the possible $D_{x}$ term should also be accompanied by a weak symmetric anisotropy term on the same bond:
	$$\mathcal{\hat{H}}''=\delta_{x}\sum\limits_{j}\spin^{x}_{j,1}\spin^{x}_{j+1,1}+\spin^{x}_{j,2}\spin^{x}_{j+1,2}$$
	This is another possible source of non-vanishing gap at the ``compensation field'' $B^{\ast}$. After the transformation it becomes:
	\begin{equation}\label{eqn:KSEA_XXZ}
	\mathcal{\hat{H}}_{\text{XXZ}}''=\delta_{x}\sum\limits_{j}\hat{T}^{x}_{j}\hat{T}^{x}_{j+1}.
	\end{equation}
	
	Actually, both $\mathcal{\hat{H}}_{\mathrm{XXZ}}'$ and $\mathcal{\hat{H}}_{\mathrm{XXZ}}''$ yield effectively the same relevant interaction responsible for the generation of the gap~\cite{SM}. We have checked that $\delta_{z}$ component of KSEA generated by $D_{z}$ does not contribute to the gap since it does not break the U(1) symmetry around the field direction $z$.
	The effective spin chain Hamiltonian (\eqref{EQ:HamiltonianXXZ}, \eqref{eq:Hamilt_DM_XXZ}, \eqref{eqn:KSEA_XXZ}) can be bosonized using the standard methods \cite{Giamarchi_1D,Bouillot_PRB_2011_bigBPCBpaper,SM}. This yields the effective low-energy field theory near $B\approx B^\ast$ in the form of sine-Gordon-type Hamiltonian,
	\begin{equation}
	\hat{\mathcal H}_{\rm eff} =  \frac{v}{2\pi} \int dx \biggl[K\biggl(\frac{\partial \phi}{\partial x}\biggr)^2 + \frac 1K \biggl(\frac{\partial \theta}{\partial x} \biggr)^2 \biggr]  + \lambda \int dx \cos(2\theta).
	\label{eqn:sg}
	\end{equation}
	$\phi$ and $\theta$ are bosonic fields noncommutative with each other.
	$K$ represents the strength of interactions and is called the Luttinger parameter~\cite{Giamarchi_1D}.
	The coupling constant $\lambda$ is proportional to $(D_x/J_\parallel)^2$ for the following reason.
	The DM interaction \eqref{eq:Hamilt_DM_XXZ} directly yields a complex interaction, $\cos \theta \sin(2\phi)$.
	This interaction itself is negligible in the low-energy Hamiltonian but it generates the excitation gap indirectly by yielding the relevant interaction $\cos(2\theta)$ through a second-order perturbative process to the TLSL~\cite{SM}.
	This indirect generation of $\cos(2\theta)$ can be regarded as the effective generation of the symmetric exchange anisotropy \eqref{eqn:KSEA_XXZ}~\cite{Furuya_BPCB} by the DM interaction \eqref{eq:Hamilt_DM_XXZ}.

	The cosine interaction potential $\cos(2\theta)$ pins the $\theta$ field to one of its minima in the ground state.
	The pinning is accompanied by a spontaneous breaking of the translation symmetry $\hat{\bm{T}}_j\to\hat{\bm{T}}_{j+1}$.
	One of the ground state has a transverse staggered magnetization $\langle\sum_j(-1)^j\hat{T}_j^x\rangle >0$ and the other has $\langle\sum_j(-1)^j\hat{T}_j^x\rangle <0$.
	Once the system chooses one of the doubly degenerate ground states spontaneously,  its effective field theory \eqref{eqn:sg} is reduced to be that of quantum spin chains in a transverse staggered field~\cite{Oshikawa_ESR_2002, Zvyagin_CuPM, SM, Kuzmenko, Furuya_SG}.

	Magnetic excitations generated by $\phi$ and $\theta$ fields were closely investigated in the context of ESR~\cite{Oshikawa_ESR_2002, Furuya_SG}.
	Applying those ESR theories to BPCB, we conclude that our ESR measurements captured a single soliton (or an antisoliton) excitations at low temperatures.
	The soliton and the antisoltion are topological excitations which cause a tunneling of $\theta$ from one minimum of the cosine potential to another one~\cite{Faure_bacovo}.
	In a magnetic field $B\approx B^\ast$, the pseudospin is bosonized as,
	\begin{equation}
	\hat{T}_j^+=e^{-i\theta}\biggl[(-1)^jB^\ast+b_1\cos\biggl(2\phi+2Kg\mu_B(B-B^\ast)\frac{x}{v\hbar}\biggr)\biggr],
	\end{equation}
	where $B^\ast$ and $b_1$ are nonuniversal constants~\cite{Giamarchi_1D, Hikihara_bosonization}.
	The $\theta$ field is pinned to a constant and an operator $\exp(2i\phi(x))$ generates a soliton at a position $x$.
	ESR thus detects a delta-function-like peak at a frequency~\cite{Furuya_Momoi}:
	\begin{equation}
	2\pi\hbar\nu=\sqrt{M^2+(2Kg\mu_B(B-B^\ast))^2},
	\end{equation}
	which corresponds to an excitation of a soliton \cite{Zamolodchikov, Lukyanov} with the mass $M$, equal to the observed gap $\Delta$, at an incommensurate wave number $q=2Kg\mu_B(B-B^\ast)/(\hbar v)$ along the chain.

	The effective $g$-factor thus turns out to be
	\begin{equation}\label{eqn:g_eff}
	g_{\rm eff} = 2Kg.
	\end{equation}
	Within the pseudospin approximation~\eqref{EQ:HamiltonianXXZ} the Luttinger parameter $K=3/4$ at $B=B^\ast$ ~\cite{Giamarchi_1D, Cabra}.
	The theoretical prediction \eqref{eqn:g_eff} is consistent with the observed anomalously large $g$ factor including its field-orientation dependence. Taking into account higher orders of $J_{\parallel}/J_{\perp}$ expansion \cite{Bouillot_PRB_2011_bigBPCBpaper} one obtains for BPCB $K=0.8$ (see details in ~\cite{SM}), which yields $g$-factor values close to the experimentally observed one (see Table~\ref{tab:g_factor}).

	\begin{table}
		\caption{Comparison of experimentally measured $g$-factors with the predictions of TLSL model.}\label{tab:g_factor}
		\begin{ruledtabular}
			\begin{tabular}{cccc}
				&\multicolumn{2}{c}{Experiment}&TLSL theory\\
				~& $g~(T>10$~K)& $g_{\text{eff}}$ ($T=0.45$~K) & (Eqn.\eqref{eqn:g_eff})\\
				\hline
				$B\parallel b$ & 2.18 $\pm$ 0.01 & 3.32 $\pm$ 0.05 & 3.49\\
				$B\parallel N$ & 2.29 $\pm$ 0.02 & 3.67 $\pm$ 0.07 & 3.66\\
				$B\parallel N$ & 2.04 $\pm$ 0.02 & 3.40 $\pm$ 0.05 & 3.26\\
				
			\end{tabular}
		\end{ruledtabular}
	\end{table}

	Unfortunately, it remains challenging to give a microscopic explanation to the field-orientation dependence of the soliton gap.
	This is because no microscopic information is available yet about the nonuniversal proportionality coefficient between the coupling constant $\lambda$ and squares of DM vectors in general field orientations~\cite{SM}.
	Still, one can ascertain that at $B \parallel N$ the soliton gap is decreasing as $|B-B^\ast|$ increases.
	This field dependence explains the reason why we observe the gapped soliton mode in ESR only in the small field range around $B= B^\ast$.
	
	We note that breather modes, which are bound states of the soliton and the antisoliton, are formed but invisible in ESR experiment because of a mismatch of wave numbers.
	The breather modes are developed near the wave number $q=\pi$ whereas ESR sees excitations at $q\approx0$.
	A staggered DM interaction is requisite for rendering breathers observable in ESR~\cite{Oshikawa_ESR_2002, Furuya_SG} but it is forbidden in BPCB by the symmetry.

	We also can estimate possible DM vector orientation from soliton gap values. From the bosonization theory, we know that $\Delta \propto {(D_{\perp})}^{\frac{4K}{4K-1}}$, where $K$ is the Luttinger parameter and $D_{\perp}$ is DM magnitude transverse to the magnetic field. DM vector directions in inequivalent ladders are linked by crystall symmetry. By taking ratios of ESR gaps and assuming $K=0.8$ as follows from the TLSL model we obtained DM vector directions (see Fig. \ref{fig:Geom01}). Found projections of DM vectors on $(bc)$ plane are practically the same as in Ref.~\onlinecite{Zvyagin_BPCB}, but our analysis predicts that DM vector component parallel to the ladder has approximately the same length as the component transverse to the ladder. Procedure of estimation is described in details in the Supplemental Materials ~\cite{SM}.

	We also put special emphasis on the coexistence of the gapless TLSL behavior and the gapped soliton mode.
	This coexistence originates from differences of two types of spin ladders.
	The spin ladder has the finite soliton gap in a field range $B^\ast-\delta B \le B \le B^\ast + \delta B$, where $\delta B$ corresponds to the soliton gap at $h_{\rm eff}=0$ through $\delta B = M/(g_{\rm eff}\mu_B)$~\cite{SM}.
	We estimate from our experimental data that the first ladder has the soliton gap for $9.32~\mathrm{T} \le B \le 10.2~\mathrm{T}$ and the second ladder has the gap for $10.1~\mathrm{T} \le B \le 10.7~\mathrm{T}$.
	When the first ladder has the gap, the second ladder remains gapless in the TLSL phase, and vice versa.
	Both spin ladders are gapped simultaneously, if they could, in an invisibly narrow field range.

	To summarize, this Letter illustrates how crucial could be presence of anisotropy in the system, it leads to symmetry breaking and ground state changes. Gapped behavior of frequency-field dependencies indicates presence of massive modes at $q=0$ in region where massless modes were expected.
	The DM vector that breaks the U(1) spin-rotation symmetry is responsible for the gapped mode.
	Our analysis with the DM vector is not only consistent with a previous ESR theory of BPCB~\cite{Furuya_BPCB} but also provides more realistic microscopic model of BPCB.
	This untypical and unexpected behavior of energy spectra bridges theory and experiment and provides one more example of quantitative test of the TLSL model, leading to the deeper understanding of low-dimensional systems physics.
	
	\acknowledgments
	
	We thank Prof. A. I. Smirnov, Prof. L. E. Svistov and T. A. Soldatov for helpful discussions. Samples were grown by evaporation method in ETH Zurich.
	ESR experiments were performed in P. L. Kapitza Institute.

	The experimental work was supported by Russian Science Foundation Grant No. 17-12-01505.
	Data analysis was supported by Program of fundamental studies of HSE. Work at ETHZ was supported by Swiss National Science Foundation, Division II.

	
	
	%
\end{document}


\title{Supplementary material: Anisotropy-induced soliton excitation in the magnetized strong-rung spin ladder}
	
	\author{Yu. V. Krasnikova}
	\affiliation{P. L. Kapitza Institute for Physical Problems, RAS, Kosygina 2, 119334 Moscow, Russia}
	\email{krasnikova.mipt@gmail.com}

	\affiliation{Laboratory for Condensed Matter Physics, National Research University Higher School of Economics, Myasnitskaya str.20, 101000 Moscow, Russia}
	
	\author{S. C. Furuya}
	\affiliation{Condensed Matter Theory Laboratory, RIKEN, Wako, Saitama 351-0198, Japan}
	
	\author{V. N. Glazkov}
	\affiliation{P. L. Kapitza Institute for Physical Problems, RAS, Kosygina 2, 119334 Moscow, Russia}
	
	\affiliation{Laboratory for Condensed Matter Physics, National Research University Higher School of Economics, Myasnitskaya str.20, 101000 Moscow, Russia}
	
	\author{K. Yu. Povarov}
	\affiliation{Laboratory for Solid State Physics, ETH Z\"{u}rich, 8093 Z\"{u}rich, Switzerland}
	
	\author{D. Blosser}
	\affiliation{Laboratory for Solid State Physics, ETH Z\"{u}rich, 8093 Z\"{u}rich, Switzerland}
	
	\author{A. Zheludev}
	\affiliation{Laboratory for Solid State Physics, ETH Z\"{u}rich, 8093 Z\"{u}rich, Switzerland}
	
	\date{\today}
	
	\maketitle

	\section{XXZ mapping}
	Here, we describe an approximate mapping from the strong-rung spin ladder into an XXZ spin chain.
	In initial approximation BPCB in the magnetic field is well-described by a Heisenberg spin ladder model with Zeeman term,
	\begin{equation}
	\mathcal{H}=J_{\perp}\sum\limits_{i=j}\vect{S}_{j,1}\cdot\vect{S}_{j,2}+J_{\parallel}\sum\limits_{n=1,2}\sum\limits_{j}\vect{S}_{j+1,n}-g\mu_{B}\sum\limits_{n=1,2}\sum\limits_{j}BS_{j,n}^z,
	\end{equation}
		When the ladder has a significant rung exchange coupling ($J_{\perp} \gg J_{\parallel}$), the spin ladder system at low energies can be described as a system of weakly coupled dimers. 
		In the magnetic field $(B>B_{c1})$, one antiferromagnetic dimer can be seen as a  two-level energy system of 1/2 pseudospin $\vect{T}_j$.	
		Within this approximation, the ladder can be mapped onto the XXZ chain by the following transformation,
		\begin{equation}
		S_{j,n}^{\pm}=\frac{(-1)^{n}}{\sqrt{2}}T^{\pm}_{j}, \quad S_{j,n}^z=\frac{1}{4}(1+2T_{j}^{z})
		\end{equation}
		The Hamiltonian of the pseudospin chain is written as,
		\begin{equation}
		\mathcal{H}_\text{eff}=J_{\perp}(T_{j}^{x}T_{j+1}^{x}+T_{j}^{y}T_{j+1}^{y}+\Delta_{z}T_{j}^{z}T_{j+1}^{z})-h_\text{eff}\sum\limits_{j}T_{j}^{z},
		\end{equation}
		where the uniaxial anisotropy parameter $\Delta_z$ and the effective field $h_{\rm eff}$ are $\Delta_{z}=1/2, h_\text{eff}=g\mu_{B}B-J_{\perp}-\frac{1}{2}J_{\parallel}=g\mu_{B}(B-B^{*})$.
		The symmetry of BPCB allows Dzyaloshinskii-Moriya interactions which is uniform on legs and forbidden on rungs,		
		\begin{equation}
		\mathcal{H}'=\sum\limits_{n=1,2}\sum\limits_{j}(-1)^n \vect{D} \cdot [\vect{S}_{j,n} \times \vect{S}_{j+1,n}], 
		\end{equation}		
		Therefore, if we assume $\vect D=(D_{x}, 0 , D_{z})$, we can rewrite the pseudospin Hamiltonian as,		
		\begin{equation}
		\mathcal{H}_\text{eff}=J_{\parallel}\sum\limits_{j}(T_{j}^{x}T_{j+1}^{x}+T_{j}^{y}T_{j+1}^{y}+\frac{1}{2}T_{j}^{z}T_{j+1}^{z})-h_\text{eff}\sum\limits_{j}T_{j}^{z}+\frac{D_{x}}{\sqrt{2}}\sum\limits_{j}(T_{j}^{y}T_{j+1}^{z}-T_{j}^{z}T_{j+1}^{y}).
		\end{equation}
	The 1D XXZ chain can be described in Tomonaga-Luttinger spin liquid model. Luttinger parameter $K$ with $\Delta_{z}$ for the pseudospin XXZ chain is \cite{Cabra, Giamarchi_1D}:	
	\begin{equation}
	K=\frac{\pi}{2(\pi-\cos^{-1}{\Delta_{z}})}=\frac{3}{4}.
	\end{equation}

	\section{Bosonization}
	
	We can bosonize the pseudospin chain using the formula,
	\begin{equation}
	\vect{T}_{j}=\vect{J}+(-1)^{j}\vect{N},
	\end{equation}
	with
	\begin{align}
	J^{z}=\frac{a_0}{\pi}\partial_{x}\phi, \\
	N^{z}=a_1\cos(2\phi), \\
	N^{+}=b_0 e^{i\theta}, \\
	J^{+}=b_1 e^{i\theta}\cos(2\phi).
	\end{align}
	The bosonized Hamiltonian is written as,
	\begin{equation}
	\mathcal{H}_{\text{eff}}=\frac{v}{2\pi}\int dx \Bigr( K(\partial_x \theta)^2 + \frac{1}{K}(\partial_x \phi)^2 \Bigl)-\frac{h_{\text{eff}}}{\pi} \int dx \partial_x\phi+cD_{x}\int dx\cos{\theta}\sin(2\phi)
	\label{H_bare}
	\end{equation}
	where $c$ is a real constant.
	The last trigonometric term is derived from the following operator product expansion.
	\begin{align}
	\partial_x \phi(x+a_0) \cos(2\phi(x)) = \frac{i\sqrt{K}}{2a_0} \sin(2\phi(x)) + \cdots,
	\end{align}	
	where less irrelevant interactions are discarded.
	This operator-product expansion leads to
	\begin{align}
	\frac{D_x}{\sqrt 2} \int dx \, (J^y(x) J^z(x+a_0) - J^z(x) J^y(x+a_0)) 
	&\approx -\frac{D_x b_1}{\pi}\sqrt{\frac K2} \int dx \, \cos \theta(x) \sin (2\phi(x)) + \cdots.
	\end{align}
	The staggered component $N^a(x)$ of the spin operator yields similar but much less relevant interactions.
	The coefficient $c\approx -(b_1/\pi)\sqrt{K/2}$ is a nonuniversal constant its precise value depends on details of the lattice model.
	The term $\cos{\theta}\sin(2\phi)$ does not yield directly the excitation gap. Still, it can yield the excitation gap indirectly by generating a relevant interaction $\cos(2\theta)$. 
	The interaction $\cos(2\theta)$ is effectively generated through the renormalization-group transformation.
	Starting from the Hamiltonian \eqref{H_bare}, we perform the renormalization group transformation repeatedly to obtain the low-energy one.
	The low-energy Hamiltonian contains several interactions in addition to those in Eq.~\eqref{H_bare}:
	\begin{align}
	\mathcal H_{\rm eff} &=\frac{v}{2\pi}\int dx \Bigr( K(\partial_x \theta)^2 + \frac{1}{K}(\partial_x \phi)^2 \Bigl)-\frac{h_{\text{eff}}}{\pi} \int dx \partial_x\phi
	\notag \\
	&\qquad + g_1\int dx\cos{\theta}\sin(2\phi) + g_2 \int dx \cos (2\theta) + g_3 \int dx \cos(4\phi).
	\end{align}
	The bare values of the coupling  constants $g_2$ and $g_3$ are zero.
	However, they become nonzero in the course of the renormalization-group transformation.
	In an early step of the renormalization-group transformation,  the coupling constants $g_2$ and $g_3$ are proportional to $(g_1)^2 = (cD_x)^2$ because the operator-product expansion of $\cos \theta \sin 2\phi$ with itself generates $\cos(2\theta)$:
	\begin{align}
	 \cos\theta(x) \sin(2\phi(x)) \cos \theta(y) \sin(2\phi(y)) 
	 &=\frac 1{4|x-y|^{2K}} \cos\theta(x) \cos \theta(y) -\frac 1{4|x-y|^{\frac 1{2K}} }\sin(2\phi(x)) \sin(2\phi(y)) + \cdots
	 \notag \\
	 &= \frac 1{8|x-y|^{2K-\frac 1{2K}}} \cos (2\theta(y)) + \frac 1{8|x-y|^{\frac 1{2K}-2K}} \cos(4\phi(y)) +\cdots.
	\end{align}
	This operator-product expansion leads to the following perturbative renormalization-group equations,
	\begin{align}
	\frac{dg_1(\ell)}{d\ell} &= \biggl(2-\frac 1{4K}- K \biggr) g_1(\ell) + \frac 18 g_1(\ell) \{ g_2(\ell) + g_3(\ell)\}, \\
	\frac{dg_2(\ell)}{d\ell} &= \biggl(2-\frac 1K \biggr) g_2(\ell) +\frac 1{16} \bigl(g_1(\ell)\bigr)^2, \\
	\frac{dg_3(\ell)}{d\ell} &= (2-4K) g_3(\ell) +\frac 1{16} \bigl( g_1(\ell)\bigr)^2.
	\end{align}
	Here, $\ell = \ln (a_0/a)$ is a parameter that relates the effective short-range cutoff $a$ with its bare value $a_0$.
	Because $\max\{\frac 1{K}, \frac 1{4K} + K\}< 2 < 4K$ for $h_{\rm eff} \approx 0$, two coupling constants $g_1$ and $g_2$ grow in the low-energy limit but the other one $g_3$ vanishes there.
	Hence, the $\cos(4\phi)$ term can be ignored in the low-energy limit.
	On the other hand, the $\cos \theta \cos(2\phi)$ term is also ignored despite its relevance because the following reason.
	When the Tomonaga-Luttinger liquid acquires the excitation gap, either $\theta$ or $\phi$ is pinned to a certain constant.
	Because of the commutation relation $[\phi(x), \partial_y(y)] =i\pi \delta(y-x)$, when the $\theta$ field is pinned, $\phi$ becomes extremely uncertain and vice versa.
	Such an interaction only affects the high-energy physics which is out of scope of our study.
	Therefore,  we can take $g_1=g_3=0$ in the low-energy Hamiltonian.

	We note that can regard the effective generation of $\cos(2\theta)$ as the generation of the KSEA interaction by the DM interaction \cite{Kaplan_ZPB_1983_KSEA1,*Shekhtman_PRL_1992_KSEA2,*Shekhtman_PRB_1993_KSEA3},	
	\begin{equation}
	\mathcal{H}'_{\text{KSEA}}=\delta_{x} \sum\limits_{j}T_{j}^{x}T_{j+1}^{x}.
	\end{equation}	
	We have checked that $\delta_{z}$ component of the KSEA interaction generated by $D_{z}$ does not contribute to the gap since it does not break the U(1) symmetry around the field direction $z$. 
	
	Taking into account the effective generation of relevant KSEA interaction and discarding the irrelevant interactions, we can rewrite low-energy effective Hamiltonian as	
	\begin{equation}
	\mathcal{H}_{\text{eff}} \approx \frac{v}{2\pi} \int dx \Bigr(K(\partial_{x}\theta)^2+\frac{1}{K}(\partial_{x}\phi)^2\Bigl)-\frac{h_{\text{eff}}}{\pi} \int dx \partial_x\phi+ \lambda \int dx \cos(2\theta),
	\end{equation}
	with $\lambda=g_2$ is a function of $(D_x)^2$. The excitation gap in the limit of $K \rightarrow 1/2$ was discussed in Ref. \cite{Hikihara_bosonization}. Compared to the $K \rightarrow 1/2$ case, the $\cos(2\theta)$ is more relevant in our case since $K \simeq 3/4$.
	Shifting $\phi(x) \rightarrow \phi(x)+\frac{Kh_{\text{eff}}}{v}x$, we can further simplify the Hamiltonian to that of the sine-Gordon theory,
	\begin{equation}
	\mathcal{H}_{\text{eff}}=\frac{v}{2\pi} \int dx \Bigr(K(\partial_{x}\theta)^2+\frac{1}{K}(\partial_{x}\phi)^2\Bigl)+ \lambda \int dx \cos(2\theta).
	\end{equation}
	The ground state of this sine-Gordon theory is doubly degenerate. The ground states break the translation symmetry spontaneously and have the expectation value $\langle \cos\theta \rangle > 0$ or $\langle \cos\theta \rangle < 0$. In other words, the ground state has either $\langle m_s \rangle >0$ or $\langle m_s \rangle <0$, where 
	\begin{equation}
	m_{s}=\sum\limits_{j}(-1)^j T_{j}^{x},
	\end{equation}
	is the transverse N\'eel order of the pseudospin.	
	Once one of the ground states is chosen spontaneously, the excitation above the ground state is described by another sine-Gordon theory,	
	\begin{equation}
	\mathcal{H}_{\text{eff}}=\frac{v}{2\pi} \int dx \Bigl(K(\partial_{x}\theta)^2+\frac{1}{K}(\partial_{x}\phi)^2\Bigr)+\tilde \lambda \int dx \cos(\theta),
	\end{equation}	
	with $\tilde \lambda \propto 2 \lambda m_{s}$.
	Here, we performed a mean-field approximation and replaced $\cos(2\theta)$  to $2m_{s}\cos\theta$. This replacement corresponds to the mean-field approximation $(T_{j}^{x})^2 \approx (-1)^{j}2m_{s}T_{j}^{x}$ to the effective KSEA interaction.	
	Elementary excitations of the sine-Gordon theory are the soliton and the antisoliton which have the degenerate excitation gap $\Delta$. The excitation gap is given by \cite{Zamolodchikov}:
	\begin{equation}
	\Delta=\frac{v}{2a_{0}\sqrt{\pi}}\frac{\Gamma(\frac{\xi}{2})}{\Gamma(\frac{1+\xi}{2})} \left( \frac{a_0\pi}{2}\frac{\Gamma(\frac{1}{1+\xi})}{\Gamma(\frac{\xi}{1+\xi})}\frac{\tilde \lambda}{v}\right)^{(1+\xi)/2},
	\end{equation}	
	where $\xi$ is the following parameter,	
	\begin{equation}
	\xi=\frac{1}{8K-1}.
	\end{equation}
	
	The Lorentz invariance of sine-Gordon theory determines the dispersion relation of the soliton and the antisoliton:
	
	\begin{equation}
	E_{s}(q)=\sqrt{\Delta^2+(vq)^2}.
	\end{equation}

	\section{ESR spectrum}
	
Let us discuss the ESR spectrum of BPCB on the basis of the bosonization analysis developed above. In the Faraday configuration with the unpolarized microwave, the ESR spectrum $I(\omega)$ as a function of frequency with a fixed magnetic field $B$ is the following (Appendix A of Ref. \cite{Furuya_Momoi}):

\begin{equation}
I(\omega) \propto -\omega[\text{Im}G_{S^+S^-}^R(\omega) +\text{Im}G_{S^-S^+}^R(\omega)],
\end{equation}	
within the linear response. $G_{S^\pm S^\mp}^{R}(\omega)$ are retarded Green's functions of the total spin $\vect{S}_{tot}=\sum\limits_{n=1,2}\sum\limits_{j}\vect{S}_{j,n}$, that is,
\begin{equation}
G_{S^\pm S^\mp}^{R}(\omega)=-i\int_{0}^{\infty}dte^{i\omega t} \langle[S^{\pm}_{\text{tot}}(t),S^{\mp}_{\text{tot}}(0)]\rangle.
\end{equation}		
	The average $\langle \cdot \rangle$ is taken with respect to the original Hamiltonian,
	\begin{equation}
	\mathcal{H}=J_{\perp} \sum\limits_{j}\vect{S}_{j,1}\cdot \vect{S}_{j,2} + J_{\parallel} \sum\limits_{n=1,2}\sum\limits_{j}\vect{S}_{j,n}\cdot \vect{S}_{j+1,n}-g\mu_{B}B\sum\limits_{n=1,2}\sum\limits_{j}\vect{S}^{z}_{j,n} + \sum\limits_{n=1,2}\sum\limits_{j}(-1)^{n}\vect{D}\cdot[\vect{S}_{j,n}\times\vect{S}_{j+1,n}].
	\end{equation}
	The total spin $S_{\text{tot}} (t)$ follows the simple equation of motion (hereafter we set $\hbar = k_{B}=1$),
	\begin{equation}
	\frac{dS^{+}_{\text{tot}}(t)}{dt}=i[\mathcal H, S^{+}_{\text{tot}}(t)]=-ig\mu_{B}BS^{+}_{\text{tot}}(t)+i\mathcal{A}(t),
	\end{equation}
	with $\mathcal{A}(t)=e^{i\mathcal{H}t}\mathcal{A}e^{-i\mathcal{H}t}$ and $\mathcal{A}=[\mathcal{H}, S^{+}_{\text{tot}}]$.
The equations of motion for $S_{\text{tot}}^{\pm}(t)$ result in a useful identity \cite{Oshikawa_Affleck},
\begin{equation}\label{EQ:Green}
G_{S^\pm S^\mp}^{R}(\omega)=\frac{2\langle S^{z}_{\text{tot}}\rangle}{\omega-g\mu_{B}B}-\frac{\langle [\mathcal{A},S_{\text{tot}}^{z}]\rangle}{(\omega-g\mu_{B}B)^2}+\frac{1}{(\omega-g\mu_{B}B)^2}G^{R}_{\mathcal{A}{\mathcal{A}}^{\dagger}}(\omega)
\end{equation}
Now we map the spin ladder to the pseudospin chain. Within the pseudospin approximation, the operators $S^{\pm}_{\text{tot}}$ of the total spin become
\begin{equation}
S^{\pm}_{\text{tot}}=\sum\limits_{n=1,2}\sum\limits_{j}S^{\pm}_{j,n}=0.
\end{equation}
Thus, if we apply the pseudospin approximation to $G_{S^\pm S^\mp}^{R}(\omega)$ directly, we would obtain $G_{S^\pm S^\mp}^{R}(\omega)=0$. However, this is not the case.
Using the identity (\ref{EQ:Green}) and applying the pseudospin approximation to its right-hand side, we obtain meaningful results to be discussed below. The ESR spectrum $I(\omega)$ within the approximation is composed of the trivial one with the resonance frequency $\omega=g\mu_{B}B$ and the additional one governed by the retarded Green's function of $\mathcal{A}$ since
\begin{equation}
-\text{Im}G^{R}_{S^+S^-(\omega)} \approx 2\pi \langle S^z_{\text{tot}} \rangle \delta(\omega-g\mu_{B}B)-\frac{1}{(\omega-g\mu_{B}B)^2}\text{Im}G^{R}_{\mathcal{A}\mathcal{A}^\dagger}(\omega)
\end{equation}
Likewise, we obtain
\begin{equation}
-\text{Im}G^{R}_{S^+S^-(\omega)} \approx 2\pi \langle S^z_{\text{tot}} \rangle \delta(\omega+g\mu_{B}B)-\frac{1}{(\omega+g\mu_{B}B)^2}\text{Im}G^{R}_{\mathcal{A}\mathcal{A}^\dagger}(\omega)
\end{equation}
The retarded Green's function $G^{R}_{\mathcal{A}\mathcal{A}^\dagger}(\omega)$ sometimes plays an important role in ESR. For example, in another spin ladder compound DIMPY, $G^{R}_{\mathcal{A}\mathcal{A}^\dagger}(\omega)$ becomes nonzero because of the uniform DM interaction and yields in additional resonance peak \cite{Ozerov}.

Within the pseudospin approximation, the $\mathcal{A}$ operator turns into
\begin{equation}
\mathcal{A}=\frac{iD_{z}}{\sqrt{2}}\sum\limits_{j}(T^{z}_{j}T^{+}_{j+1}-T^{+}_{j}T^{z}_{j+1})=-i\sqrt{2}a_1 b_0 D_z \int dx e^{i\theta} \cos \Bigl(2\phi+\frac{2Kh_{\text{eff}}}{v}x\Bigr).
\end{equation}
The operator $\exp(\pm 2i\phi)$ creates the soliton ($+$) or antisoliton ($-$). Because the ground state has $\langle \cos\theta \rangle \not = 0$, the operator $e^{i\theta}$ is basically replacable by a constant as long as we only focus on the lowest-energy excitation created by $\mathcal{A}$ and $\mathcal{A}'$. The lowest-energy excitation created by $e^{i\theta}e^{\pm2 i\phi}$ is the single soliton (or the single antisoliton). One can find arguments for the creation and the annihilation of the soliton, the antisoliton, and the breathers in Refs. \cite{Lukyanov,Kuzmenko}. Reference \cite{Furuya_Oshikawa} discusses selection rules of those excitations in the ESR spectrum.

The $\mathcal{A}$ operator gives rise to a delta-function peak,
\begin{equation}
-\text{Im}G^{R}_{\mathcal{A}\mathcal{A}^\dagger}(\omega) \propto \delta(\omega-E_{S}(2Kh_{\text{eff}}/v)),
\end{equation}
in the ESR spectrum. It is easy to see that $-\text{Im}G^{R}_{\mathcal{A}\mathcal{A}^\dagger}(\omega)$ contains the same peak. The resonance frequency of this peak is 
\begin{equation}
\omega=E_{S}(2Kh_{\text{eff}})=\sqrt{\Delta^2+(g_{\text{eff}}\mu_{B}(B-B^{*}))^2}
\end{equation}	
with the effective $g$-factor,
\begin{equation}
g_{\text{eff}}=2Kg.
\end{equation}	
$g_{\text{eff}}$ is roughly equal to 3 since $K \approx 3/4$ for $h_{\text{eff}} \approx 0$.

The XXZ anisotropy of the pseudospin chain is $\Delta_z=0.5$ because the strong-rung expansion is stopped at the first order of $J_{\parallel}/J_{\perp}$. Taking into account higher orders of this expansion \cite{Bouillot}, the XXZ anisotropy is modified to be
\begin{equation}
\Delta_z=\frac{1}{2}-\frac{3}{8}\frac{J_{\parallel}}{J_{\perp}} \approx 0.39
\end{equation}
in the case of BPCB ($J_{\perp}=$12.6K, $J_{\parallel}=$3.55 K) if we take into account the second-order term of the strong-rung expansion. Then, the Luttinger parameter becomes
\begin{equation}
K=\frac{\pi}{2[\pi-\cos^{-1}(0.39)]} \approx 0.80.
\end{equation}
It leads to $g_{\text{eff}} \approx 1.60g$ and
\begin{equation}
\begin{matrix}
g_{\text{eff}} & =
& \left\{
\begin{matrix}
3.49, & (B \parallel b) \\
3.26~\& ~3.66, & (B \parallel N).
\end{matrix} \right.
\end{matrix}
\end{equation}
These values are roughly consistent with the experimentally obtained ones.

	\section{ESR gaps}
	If we would take into account all three components of DM interaction $\textbf{D}=(D_x, D_y, D_z)$, here magnetic field is applied along $z$. Hamiltonian in mean-field approximation with effective KSEA interaction is:
	\begin{equation}
\mathcal{H}_\text{eff} \simeq \frac{v}{2\pi} \int{dx}\Bigl(K{(\partial_{x}\theta)}^2+\frac{1}{K}{(\partial_x\phi)}^2\Bigr) + {\tilde{\lambda}}\int{dx\cos{(\theta-\alpha_D)}},
	\end{equation}
	
	where $\alpha_D=\tan^{-1}(D_y/D_x)$,

	\begin{equation}
\tilde{\lambda} \propto m_s(D_{\perp}^2),
	\end{equation}
	
	$D_{\perp}$ is the part of DM transverse to the applied magnetic field.
	
	The staggered pseudospin is 
	
	\begin{equation}
m_s=\langle{\cos{(\theta-\alpha_D)}}\rangle \propto (\tilde{\lambda})^{1/(8K-1)} = (m_s D_{\perp}^2)^\frac{1}{8K-1}
	\end{equation}
	
	Thus we obtain:
	\begin{equation}
	m_s \propto (D_{\perp})^{\frac{1}{4K-1}}
	\end{equation}
	
	The soliton gap $\Delta$ depends on $D_{\perp}$ and $K$ as follows:
	\begin{equation}
	\Delta \propto {(m_s D_{\perp}^2)}^{\frac{1+\frac{1}{8K-1}}{2}}=(D_{\perp})^{\frac{4K}{4K-1}}
	\end{equation}

\begin{figure}
  \centering
  \includegraphics[width=0.8\textwidth]{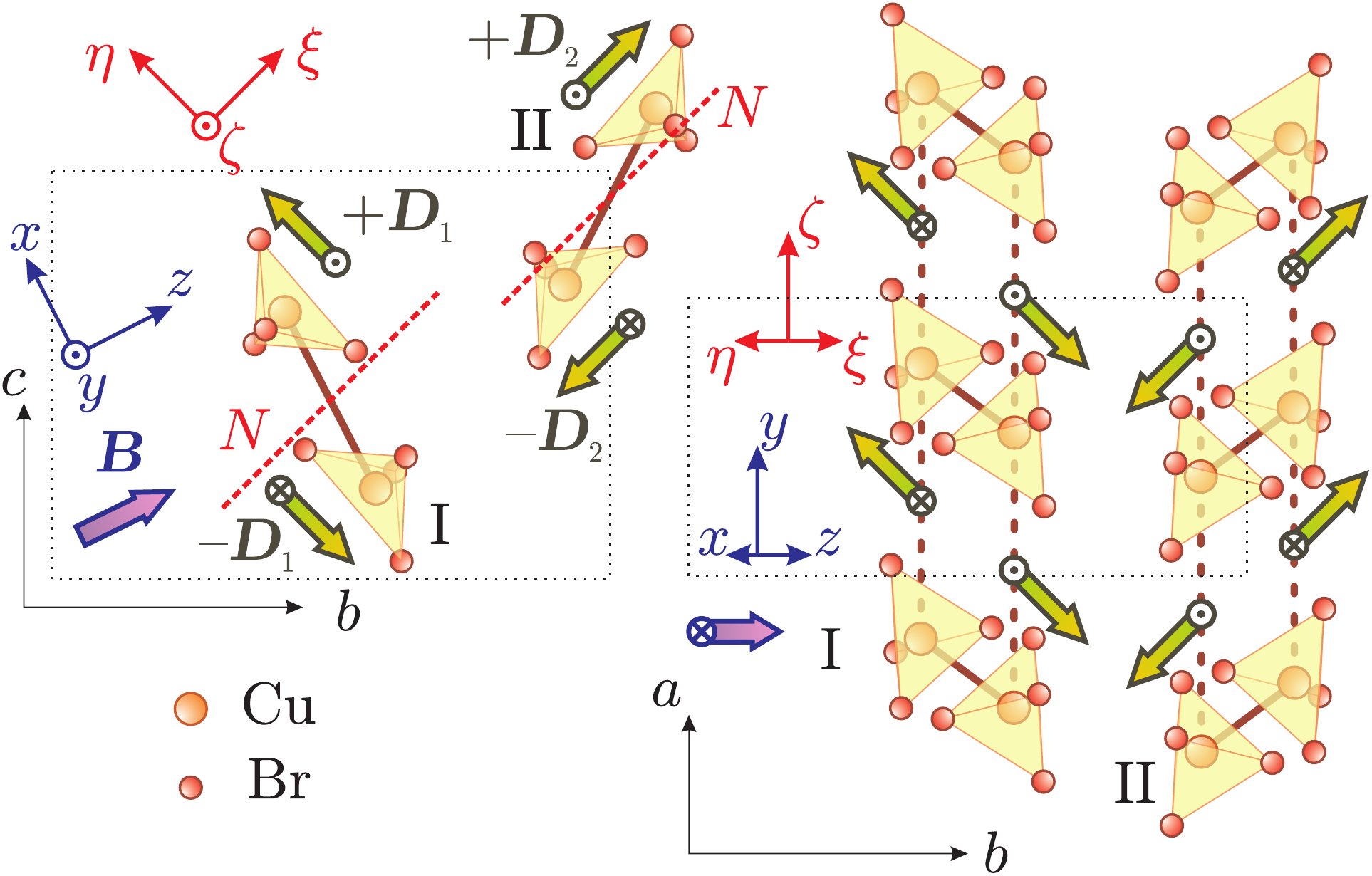}
  \caption{Geometry of DM vectors in BPCB, field (Latin) and crystal (Greek) related spin coordinate systems.}\label{FIG:Geom}
\end{figure}

\begin{figure}
	\centering	\includegraphics[width=0.8\textwidth]{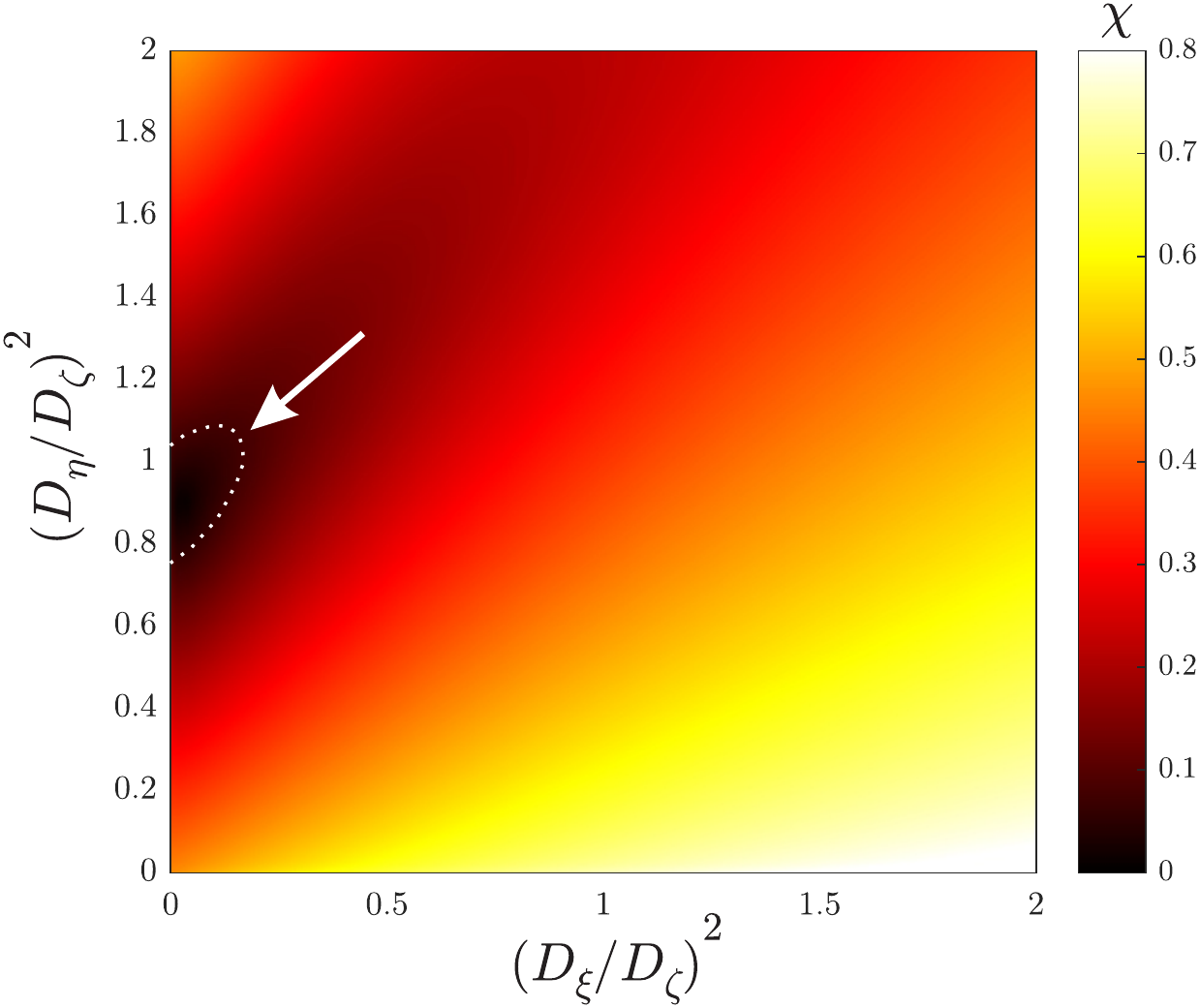}
	\caption{The criterion~(\ref{EQ:chi2}) plotted as function of $X$ and $Y$ [see definition~(\ref{EQ:simpl})] for $K=0.8$. Region with $\chi\lesssim0.1$ is highlighted on the plot. }\label{FIG:XYsurf}
\end{figure}

Symmetry of BPCB allows the following DM vectors on each of the ladder types: $\mathbf{D}_1 = (D_{\xi}, D_{\eta}, D_{\zeta})$ and $\mathbf{D}_2 = (-D_{\eta}, D_{\xi}, D_{\zeta})$. Here $\zeta \parallel a$, $\xi \parallel N$, as shown in Fig.~\ref{FIG:Geom}. Assuming that the prefactor ${A}$ is field direction independent, the gaps for  $B \parallel N$ and $B \parallel b$ are:

\begin{equation}
\Delta_{N}^{(1)}={A}(D_{\eta}^2+D_{\zeta}^2)^{\alpha/2}
\end{equation}

\begin{equation}
\Delta_{N}^{(2)}={A}(D_{\xi}^2+D_{\zeta}^2)^{\alpha/2}
\end{equation}	

\begin{equation}
\Delta_{b}={A} \Bigl(D_{\zeta}^2+\frac{(D_{\xi}+D_{\eta})^2}{2}\Bigr)^{\alpha/2},
\end{equation}	

where $\alpha = \frac{4K}{4K-1}$. In case of $B\parallel b$ the ladders are equivalent and there is just a single gap $\Delta_{b}$, while for $B\parallel N$ the gaps in the ladders I and II are $\Delta_{N}^{(1)}$ and $\Delta_{N}^{(2)}$.

It is convenient to introduce a simplified notation for the \emph{ratios} of the gaps and DM vector components:

\begin{equation}\label{EQ:simpl}
  \frac{\Delta_{N}^{(1)}}{\Delta_{N}^{(2)}}=R_{12},~\frac{\Delta_{b}}{\Delta_{N}^{(2)}}=R_{b2},~\left(\frac{D_{\xi}}{D_{\zeta}}\right)^{2}=X,~\left(\frac{D_{\eta}}{D_{\zeta}}\right)^{2}=Y.
\end{equation}

In this notation:

\begin{align}\label{EQ:relationsZ2}
  R_{12}& =\left(\dfrac{1+Y}{1+X}\right)^{\alpha/2}  \\
  R_{b2} & =\left(\dfrac{1+\frac{1}{2}\left(\sqrt{X}+\sqrt{Y}\right)^{2}}{1+X}\right)^{\alpha/2}
\end{align}

Relations~(\ref{EQ:relationsZ2}) are the important consequence of 'solitonic' excitation picture. They link the direction of DM vector in BPCB to the experimentally observed gap ratios.

For a given $K=0.8$ that follows from TLSL calculations we need to find the non-negative parameters $X$ and $Y$ that satisfy relations~(\ref{EQ:relationsZ2}). Thus, one can say that we experimentally determine the \emph{orientation} of DM vector. We can define the deviation as:

\begin{equation}\label{EQ:chi2}
  \chi=\sqrt{\left(\dfrac{R_{12}^{\mathrm{Obs}}-R_{12}^{\mathrm{Calc}}}{R_{12}^{\mathrm{Obs}}}\right)^{2}+\left(\dfrac{R_{b2}^{\mathrm{Obs}}-R_{b2}^{\mathrm{Calc}}}{R_{b2}^{\mathrm{Obs}}}\right)^{2}}
\end{equation}

As Fig.~\ref{FIG:XYsurf} shows, for $K=0.8$ the experimental gap ratio is reproduced for $D_{\xi}\simeq0$ and $D_{\eta}\simeq 0.9 D_{\zeta}$. Found projections of DM vectors on $(bc)$ plane are the same as in Ref.~\onlinecite{Zvyagin_BPCB}, but our analysis predicts that DM vector component parallel to the ladder has approximately the same length as the component transverse to the ladder. The resulting orientations of DM vectors in the ladders are shown in Fig.~\ref{FIG:Geom}.

\bibliography{supplBPCBESR}